\documentclass[a4paper,prl,twocolumn,showpacs,superscriptaddress,floatfix,nofootinbib,reprint]{revtex4-1}
\usepackage{lipsum}
\usepackage{graphicx}
\usepackage{epsf}
\usepackage{epsfig}
\usepackage{amssymb,amsmath,amsfonts}
\usepackage[usenames]{color}
\usepackage{amssymb}
\usepackage{times}
\usepackage{bm}
\usepackage{hyperref}
\hypersetup{
  colorlinks=true,        
  linkcolor=blue,         
  citecolor=cyan,         
}

\usepackage{graphicx}  
\usepackage{dcolumn}   
\usepackage{bm}        
\usepackage{amsfonts,amsmath,amssymb,mathrsfs}
\usepackage{color}

\usepackage{hyperref}
\hypersetup{
  colorlinks=true,        
  linkcolor=blue,         
  citecolor=cyan,         
}
\usepackage{natbib,times}
\definecolor{orange}{rgb}{1,0.5,0}

\newcommand{\cf}{cf.~}
\newcommand{\ie}{i.e.,~}
\newcommand{\eg}{e.g.,~}

\renewcommand{\BibitemShut}[1]{}

\newcommand{\ITP}{Institut f{\"u}r Theoretische Physik,
  Max-von-Laue-Stra{\ss}e 1, 60438 Frankfurt, Germany}
\newcommand{\FIAS}{Frankfurt Institute for Advanced Studies,
  Ruth-Moufang-Stra{\ss}e 1, 60438 Frankfurt, Germany}

\begin{document}

\title{New constraints on radii and tidal deformabilities of neutron
  stars from GW170817}
\author{Elias~R.~Most}
\affiliation{\ITP}
\author{Lukas~R.~Weih}
\affiliation{\ITP}
\author{Luciano~Rezzolla}
\affiliation{\ITP}
\affiliation{\FIAS}
\author{J\"urgen~Schaffner-Bielich}
\affiliation{\ITP}

\begin{abstract}
  We explore in a parameterized manner a very large range of physically
  plausible equations of state (EOSs) for compact stars for matter that
  is either purely hadronic or that exhibits a phase transition. In
  particular, we produce two classes of EOSs with and without phase
  transitions, each containing one million EOSs. We then impose
  constraints on the maximum mass, ($M < 2.16 M_{\odot}$), and on the
  dimensionless tidal deformability ($\tilde{\Lambda} <800$) deduced from
  GW170817, together with recent suggestions of lower limits on
  $\tilde{\Lambda}$. Exploiting more than $10^9$ equilibrium models for each
  class of EOSs, we produce distribution functions of all the stellar
  properties and determine, among other quantities, the radius that is
  statistically most probable for any value of the stellar mass. In this
  way, we deduce that the radius of a purely hadronic neutron star with a
  representative mass of $1.4\,M_{\odot}$ is constrained to be
  $12.00\!<\!R_{1.4}/{\rm km}\!<\!13.45$ at a $2$-$\sigma$ confidence
  level, with a most likely value of $\bar{R}_{1.4}=12.39\,{\rm km}$;
  similarly, the smallest dimensionless tidal deformability is
  $\tilde{\Lambda}_{1.4}\!>\!375$, again at a $2$-$\sigma$ level. On the
  other hand, because EOSs with a phase transition allow for very compact
  stars on the so-called ``twin-star'' branch, small radii are possible
  with such EOSs although not probable, \ie $8.53\!<\!R_{1.4}/{\rm
    km}\!<\!13.74$ and $\bar{R}_{1.4}=13.06\,{\rm km}$ at a $2$-$\sigma$
  level, with $\tilde{\Lambda}_{1.4}\!>\!35.5$ at a $3$-$\sigma$
  level. Finally, since these EOSs exhibit upper limits on
  $\tilde{\Lambda}$, the detection of a binary with total mass of
  $3.4\,M_{\odot}$ and $\tilde{\Lambda}_{1.7}\!>\!461$ can rule out
  twin-star solutions.
\end{abstract}


\begin{titlepage}
\maketitle
\end{titlepage}


\noindent\emph{Introduction.} On August 17 2017, the Advanced LIGO and
Virgo network of gravitational-wave detectors have recorded the signal
from the inspiral of a binary neutron-star system, \ie event GW170817
\cite{Abbott2017_etal}. Less than a couple of seconds later, the
gravitational-wave signal was followed by a series of electromagnetic
emissions. These electromagnetic counterparts have provided the
long-sought confirmation that merging neutron-star binaries can be
associated with short gamma-ray bursts, shedding important light on the
long-standing puzzle of the origin of these phenomena
\cite{Eichler89,Narayan92,Rezzolla:2011,Berger2013b}.

These multimessenger observations, together with numerical simulations of
merging neutron stars (see \cite{Baiotti2016,Paschalidis2016} for recent
reviews), and the modelling of the kilonova emission from this process
\cite{Bovard2017,Metzger2017,Perego2017} have provided important new
insight on the maximum mass of neutron stars and on the expected
distribution in radii
\cite{Annala2017,Bauswein2017b,Margalit2017,Radice2017b,Rezzolla2017,Ruiz2017,Shibata2017c}.
The approaches followed in these works differ significantly in the
techniques employed, but provide a remarkably robust picture of what is
the maximum mass of nonrotating stellar models $M_{_{\rm TOV}}$. For
example, by combining the signal from GW170817 and quasi-universal
relations (see, \eg \cite{Yagi2017,Haskell2014}) that correlate $M_{_{\rm
    TOV}}$ with the maximum mass supported through uniform rotation
$M_{\rm max}$ \cite{Breu2016} (see \cite{Weih2017} for the case of
differential rotation), Ref. \cite{Rezzolla2017} has set constraints on
the maximum mass to be $2.01^{+0.04}_{-0.04}\leq\,M_{_{\rm
    TOV}}/M_{\odot}\lesssim 2.16^{+0.17}_{-0.15}$, where the lower limit
comes from pulsar observations \cite{Antoniadis2013}. Similarly, by
considering the most generic family of neutron-star-matter equations of state (EOSs) that
interpolate between recent nuclear-physics results at low and high baryon
densities, Ref. \cite{Annala2017} has set constraints for the radius of a
$1.4\,M_{\odot}$ neutron star to be $R_{1.4}\!<\!13.6\,{\rm km}$, while
the minimum dimensionless tidal deformability is
$\tilde{\Lambda}_{1.4}\!>\!120$.

In this \emph{Letter}, we reconsider the problem of constraining the
radii and tidal deformability of neutron stars considering more than two
million different EOSs (with and without a phase transition) that are
physically plausible and respect the observational constraints on the
maximum mass. Using this large set of equilibria, we explore the
distribution functions of stellar models and how they are affected by the
imposition of various constraints, be them on the maximum mass or on the
dimensionless tidal deformability.

Explorations of this type have been considered in the recent past,
starting from the works of Refs. \cite{Ozel:2010,Steiner2010} (see also
\cite{Ozel2009}), who derived limits on the neutron-star radius by using
data from X-ray binaries combined with parametrized EOSs (see
\cite{Lattimer2016,Ozel2016} for recent reviews). When compared with
these approaches, our results benefit from several improvements. First,
we impose, and in a differential manner, recent constraints on the
maximum mass \cite{Margalit2017,Rezzolla2017, Ruiz2017,Shibata2017c} and
on the tidal deformability \cite{Perego2017,Radice2017b} coming directly
from GW170817 and that obviously could not have been included by previous
works, \eg \cite{Ozel:2010,Steiner2010,Hagen2016,Steiner2017}. Second, we
exploit recent improvements on the EOS of neutron matter in the outer
core \cite{Drischler2016}, which plays a fundamental role in determining
the stellar radius. Third, we carry out the first systematic study of the
statistical properties of the tidal deformability highlighting that the
lower limit for $\tilde{\Lambda}$ is very tightly constrained. Finally,
we explore a more recent, but also more restricted, prescription for the
outer core as infinite neutron matter \cite{Drischler2017} and compare to
the results obtained using the setup employed by
\cite{Annala2017}. Hereafter, masses will be in units of solar masses and
radii in kilometres.

\begin{figure*}
\begin{center}
  \includegraphics[width=1.8\columnwidth]{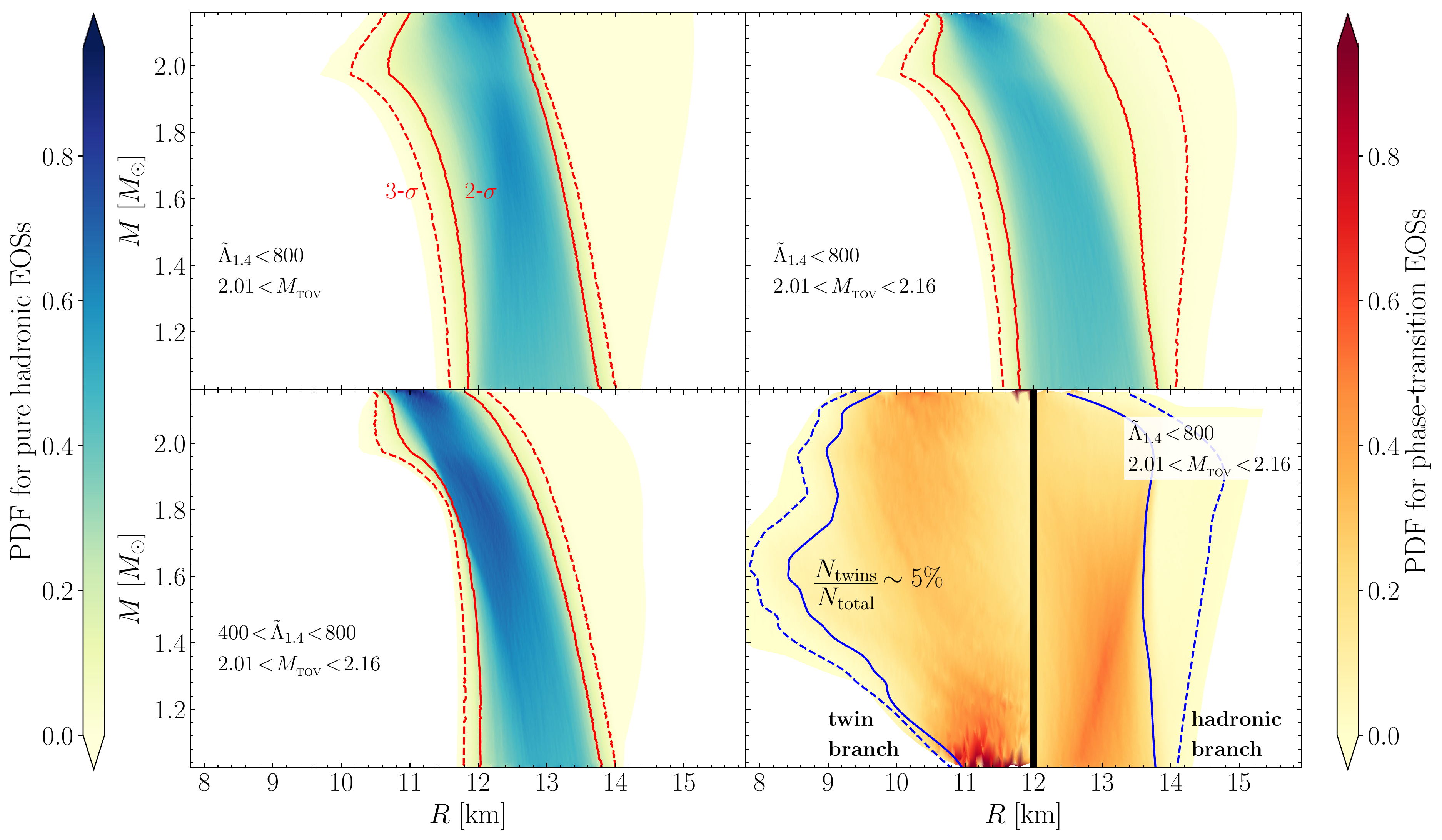}
  \caption{\footnotesize PDFs of stellar radii. Top-left panel: PDF with
    only the observational constraints on the observed maximum mass and
    tidal deformability for pure hadronic EOSs; top-right: PDF when also
    an upper limit is set on the maximum mass; bottom-left: PDF with the
    combined constraints on maximum mass and tidal deformability;
    bottom-right: the same as in the bottom-left but for EOSs with a
    phase transition; the thick black line at $12\,\rm km$ distinguishes
    the PDFs of hadronic twin stars, which represent only $5\%$ of the
    total sample with phase transitions.
    In all panels the solid and dashed lines indicate the
    $2$-$\sigma$ and $3$-$\sigma$ confidence levels, respectively.}
  \label{fig:distributions_2D}
\end{center}
\end{figure*}

\noindent\emph{Methods and setup} We compute models of cold nonrotating
neutron stars by numerically solving the Tolman-Oppenheimer-Volkoff (TOV)
equations together with an EOS. As the complete EOS is unknown, we
construct a parameterized set of EOSs by taking into account calculations
that describe nuclear matter in the outer crust \citep{
  Baym71b,Negele73}, state-of-the-art descriptions of nuclear matter
close to nuclear-saturation density \citep{Drischler2016,Drischler2017},
together with perturbative QCD calculation for matter at densities
exceeding that in the core of neutron stars
\cite{Kurkela2010,Fraga2014}. Because the EOS at intermediate densities
is not well known, we construct it using piecewise polytropes, overall
following \cite{Kurkela2014}. Additionally, we account for the existence
of phase transitions by considering EOSs that admit a jump in the energy
density between randomly chosen segments of the polytropes
\cite{Lattimer2014,Steiner2017,Tews2018a,Tews2018}, and thus allowing for
``twin-star'' solutions \cite{Alford2013,Alvarez2017,Christian2018} (see
the supplemental material for details \cite{Supplemental}).

\noindent\emph{Radius and tidal deformability constraints.} Figure
\ref{fig:distributions_2D} offers a complete view of the probability
distribution functions (PDFs) built using our $\sim2\times10^9$ stellar
models. The top-left panel, in particular, shows the colorcoded PDF when
only the \emph{observational} constraints are imposed on the maximum mass
\cite{Antoniadis2013} and on the tidal deformability
\cite{Abbott2017_etal}, \ie $2.01\!<\!M_{_{\rm TOV}}$ and
$\tilde{\Lambda}_{1.4}\!<\!800$ (see Fig. 3 of the supplemental material \cite{Supplemental}
for the PDF with only the maximum-mass constraint). Indicated with red
solid and dashed lines are the values at which the corresponding
cumulative distributions at fixed mass reach a value of $2$-$\sigma$ and
$3$-$\sigma$, respectively, thus setting both a minimum and a maximum
value for the radius at that mass with a probability of $\sim 95\%$ and
$99.7\%$. Note that the PDF extends beyond the red lines, but attains
very small values in these regions. The top-right panel shows instead the
PDF when, in addition to the lower limit, also an upper limit is set on
the maximum mass \ie $2.01\!<\!M_{_{\rm TOV}}\!<\!2.16$
\cite{Rezzolla2017}, while keeping the observational constraint on the
tidal deformability. Note that the addition of this constraint changes
the PDF, decreasing the average value of the maximum radius at a given
mass. The bottom-left panel of Fig.~\ref{fig:distributions_2D} shows the
impact of the combined observational and maximum mass constraints with
that of a lower limit on the tidal deformability as suggested by
Ref. \cite{Radice2017b}, \ie after considering $2.01\!<\!M_{_{\rm
    TOV}}\!<\!2.16$ and $400\!<\!\tilde{\Lambda}_{1.4}\!<\!800$. We note
that although the constraint $\tilde{\Lambda}_{1.4}\!>\!400$ set by
Ref. \cite{Radice2017b} does not come with a systematic quantiification
of the uncertainties, it is reasonable that such a lower limit exists on
the basis of the considerations made by Ref. \cite{Radice2017b}.
\begin{figure*}
  \begin{center}
  \includegraphics[width=0.9\textwidth]{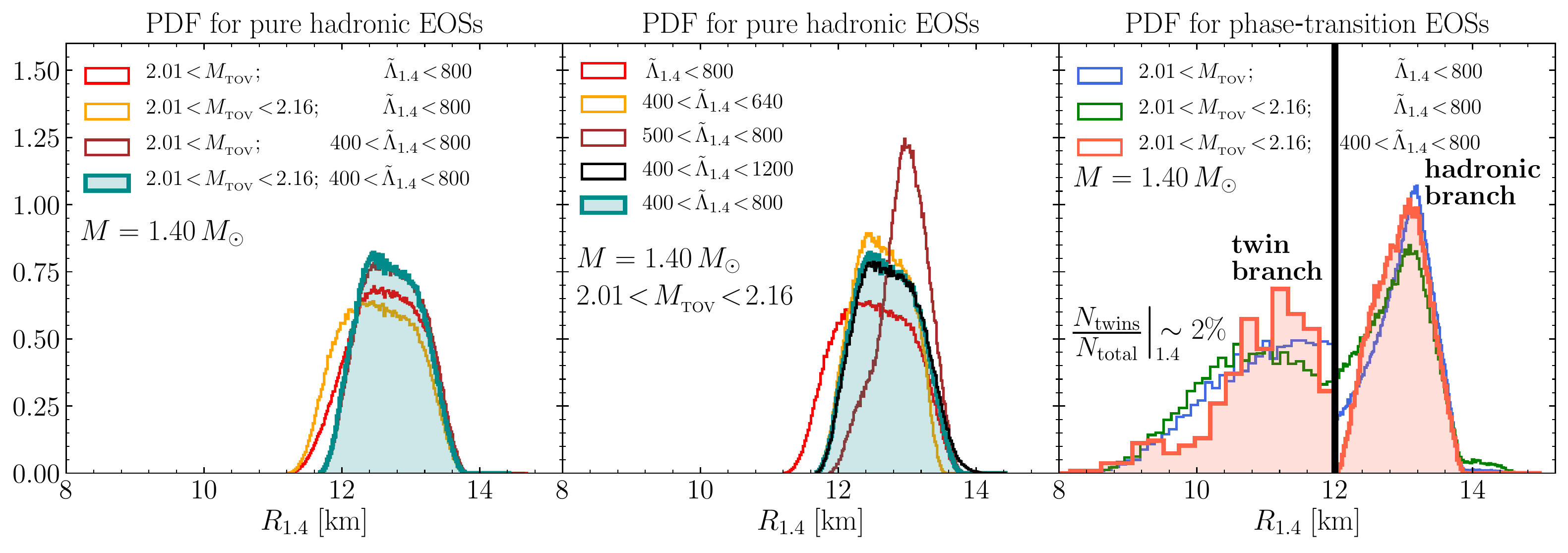}
  \caption{\footnotesize PDFs of stellar radii for a neutron star with
    mass $1.4$. Reported with different lines are the PDFs with different
    constraints on the maximum mass and tidal deformability (see
    legends); the left and middle panels refer to pure hadronic EOSs,
    while the right one to EOSs with a phase transition (\cf
    Fig. \ref{fig:distributions_2D}).}
  \label{fig:probs_M14}
  \end{center}
\end{figure*}

The effect of these combined constraints is to significantly reduce the
variance in the small-radii region and to refine the range for the most
likely radii at a given mass. Note that the distribution now is not only
restricted to a rather small range in radii, but it is also peaked around
the small-radii end of the range. Because the EOS beyond
nuclear-saturation density is not known, the possibility of phase
transitions is also taken into account in the bottom-right panel, where
we do not impose the $400\!<\!\tilde{\Lambda}_{1.4}$ constraint since it
is based on numerical simulation with EOSs without phase
transitions. Furthermore, by splitting the panel at $12\,\rm km$ we
distinguish between the PDF of the hadronic branch and the PDF of the
``twin-star'' branch, namely of all those stars that populate the
small-radii second stable branch typical of models with a phase
transition. Note that while the PDF on the hadronic branch is very
similar to the top-right panel, that of the twin-star branch is
significantly different. In particular, we find that the radius varies in
a much broader range, $8.53\!<\!R_{1.4}/{\rm km}\!<\!13.74$, and is not
as constrained as the hadronic branch; more importantly, the twin stars
represent only $\sim 5\%$ of the total sample with phase transitions.

These last results are best appreciated when considering cuts of the
bottom panels of Fig.~\ref{fig:distributions_2D} at a fixed value of the
mass, \eg $1.4$. This is shown in the left panel of
Fig.~\ref{fig:probs_M14}, which reports the PDF as a function of the
radius at that mass, $R_{1.4}$. Shown with different lines are the
distributions obtained when considering different constraints on the
maximum mass or on the tidal deformability (see legend). Note that when
only the observational constraints are imposed, either on the maximum
mass or on the tidal deformability, the distribution functions are rather
broad and flat, with a width of about almost three kilometres (\cf red
and orange lines). On the other hand, when the combined constraints are
considered, as shown with the green-shaded distribution, the variance
decreases to about two kilometres and the PDF also exhibits a peak around
the small-radii tail of the distribution. In this way, we are able to
constrain $12.00\!<\!R_{1.4}\!<\!13.45$ at a $2$-$\sigma$ confidence
level, with a most likely value of $\bar{R}_{1.4}=12.45$. Although not
shown in Fig.~\ref{fig:probs_M14}, we note that the PDFs are very robust
upon changes in upper limit of the maximum mass, both when considering
smaller ($2.1$) or larger ($2.33$) values for $M_{_{\rm
    TOV}}$. Conversely, the PDFs are rather sensitive to changes in
$\tilde{\Lambda}$. This is illustrated in the middle panel of
Fig.~\ref{fig:probs_M14}, which shows how the reference green-shaded PDF
varies when, for $M=1.4$ and $2.01\!<\!M_{_{\rm TOV}}\!<\!2.16$,
different intervals are considered for the tidal deformability (see
legend). Considering a large lower limit for the tidal deformability, \eg
from $\tilde{\Lambda}_{1.4}\!>\!400$ to $\tilde{\Lambda}_{1.4}\!>\!500$
(brown line), has the effect of excluding the softest EOSs and hence to
shift the peak of the distribution to larger values, yielding a most
likely value of $\bar{R}_{1.4}\simeq 13.0$ and a variance which is below
$2\,\rm km$. By contrast, changing the upper limit of the tidal
deformability, \eg taking the less conservative observational limit
$\tilde{\Lambda}_{1.4}\!<\!640$ \cite{Abbott2017_etal} (orange line), or
an even more conservative limit of $\tilde{\Lambda}_{1.4}\!<\!1200$
(black line), does not change the distribution significantly.
\begin{figure}[h!]
  \includegraphics[width=1.0\columnwidth]{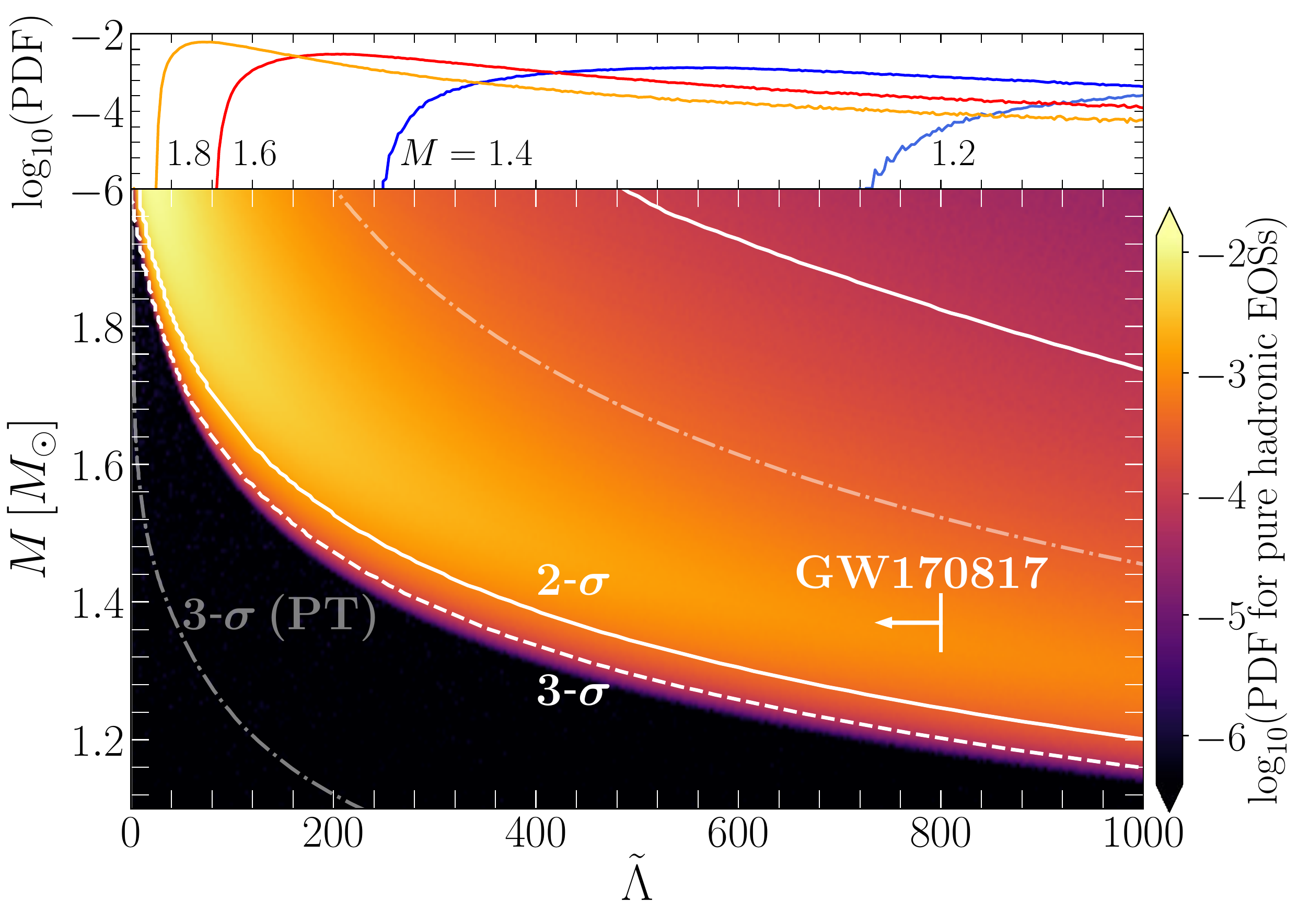}
  \caption{\footnotesize PDF of the tidal deformability $\tilde{\Lambda}$
    for pure hadronic EOSs satisfying the constraint $M_{_{\rm TOV}} >
    2.01$. The white solid and dashed lines show where the corresponding
    cumulative distributions at fixed mass reach a value of $2$-$\sigma$
    and $3$-$\sigma$, respectively. Also shown are the $3$-$\sigma$
    regions for EOS featuring a phase transition. Shown with an arrow is the upper
    limit deduced from GW170817, while several cuts at fixed masses are
    shown in the top panel.}
  \label{fig:2D_lambda}
\end{figure}
\begin{figure*}
  \includegraphics[width=1.8\columnwidth]{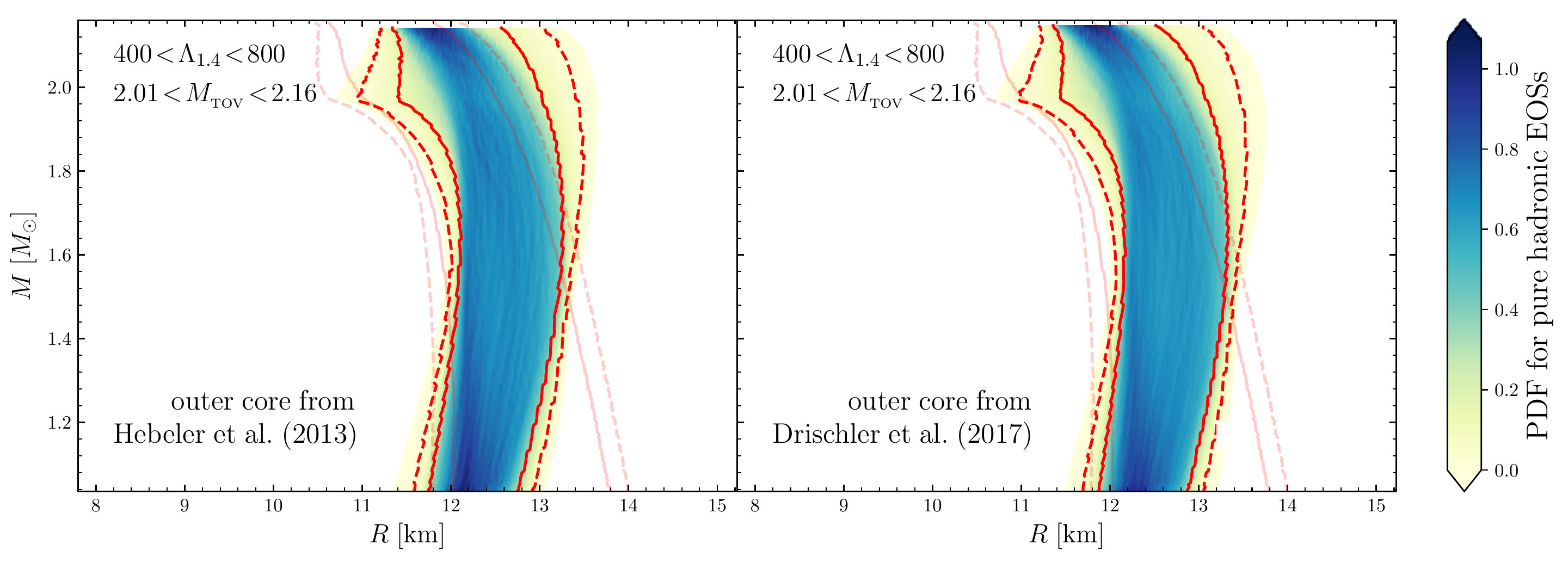}
  \caption{\footnotesize \textit{Left panel:} The same as in bottom-right
    panel of Fig.~\ref{fig:distributions_2D}, but when the neutron matter
    in the outer core is treated following the approach of
    \cite{Hebeler2013}. \textit{Right panel:} The same as in the left
    panel but when considering the more recent prescription of
    \cite{Drischler2017} for the outer core. Shown as light shaded lines
    are the $2/3$-$\sigma$ values reported in the bottom-right panel of
    Fig.~\ref{fig:distributions_2D}.}
\label{fig:new_EOSs}
\end{figure*}

Finally, when considering the distribution of models with
phase-transitions, the behaviour in the right panel of
Fig.~\ref{fig:probs_M14} is rather different. While the application of
the combined maximum mass and $\tilde{\Lambda}_{1.4}$ constraints yields
the same results presented in the left panel (green and orange shaded
curves), with $\bar{R}_{1.4}=13.06$ at a $2$-$\sigma$ level, the
twin-star branch is much broader, \ie $8.53\!<\!R_{1.4}\!<\!13.74$. Note
that although small-radii stars are possible, they are not probable and
that the twin-branch stars are only $\sim\,2\%$ of the total sample with
phase transition for $M=1.4$. Also, note that if a constraint such as
$\tilde{\Lambda}_{1.4}\!>\!400$ could be applied to EOSs with phase
transitions, it would only sharpen the mean value of the PDF for twin
stars, but again exclude the small-radii models of the hadronic branch
(orange curve).

Additional information on the tidal deformability is presented in
Fig.~\ref{fig:2D_lambda}, which reports the PDF of $\tilde{\Lambda}$ for
the hadronic EOS, with again the white solid (dashed) lines showing where
the corresponding cumulative distributions at fixed mass reach a value of
$2$-$\sigma$ ($3$-$\sigma$). Furthermore, we indicate with grey
dashed-dotted lines the $3$-$\sigma$ values for EOSs with a phase
transition (\cf Fig. 4 of the supplemental material \cite{Supplemental}). Shown instead with
an arrow is the upper limit deduced from GW170817
\cite{Abbott2017_etal}. The PDF in Fig.~\ref{fig:2D_lambda}, which has
not been presented before, points to three interesting properties. First,
at any given mass the PDF is highly asymmetrical and has a very sharp
cutoff in the lower-hand of the tidal deformability, which goes from
$\tilde{\Lambda}_{\rm cutoff}\sim 825$ at $M=1.2$ to
$\tilde{\Lambda}_{\rm cutoff}\sim 285$ at $M=1.4$ (see top panel of
Fig.~\ref{fig:2D_lambda}). Second, as the stellar mass increases, the
distribution tends to a more pronounced peak and a smaller variance, with
$0\!\lesssim\!\tilde{\Lambda}\!\lesssim\!480$ for $M=2.0$. Finally, when
considering a reference mass of $1.4$, we can set
$\tilde{\Lambda}_{1.4}\!>\!375\,(290)$, again at a $2\,(3)$-$\sigma$
confidence level, respectively. Similarly, the corresponding value for a
$1.3\,(1.5)$ mass becomes $\tilde{\Lambda}_{1.3(1.5)}\!>\!615\,(230)$ at
$2$-$\sigma$.  When allowing for phase-transitions we instead find that
at $3$-$\sigma$ $\tilde{\Lambda}_{1.4}\!>\!35.5$ and
$\tilde{\Lambda}_{1.7}\!<\!461$. Hence, a future gravitational-wave
detection of a high-mass merger with a measured value of
$\tilde{\Lambda}_{1.7}\!>\!461$ can rule out twin stars below that
mass. This is the first time that such upper limits have been provided on
twin stars (see the supplemental material \cite{Supplemental} for an extended discussion).

As a concluding but important remark, we illustrate in
Fig.~\ref{fig:new_EOSs} the impact that different treatments of the outer
core may have on the statistical properties of neutron star radii. In
particular, the left panel of Fig.~\ref{fig:new_EOSs} shows the same
distributions discussed in the bottom-left panel of
Fig.~\ref{fig:distributions_2D} when following the treatment for the
outer core discussed by Ref. \cite{Hebeler2013}, which is less
conservative than the approach used here following
Ref. \cite{Drischler2016}. Similarly, the right panel of
Fig.~\ref{fig:new_EOSs} reports the corresponding distributions when the
stellar models are built using improved neutron-matter calculations
\cite{Drischler2017}. Leaving aside the details of the two different
calculations, it is interesting that estimates for the outer core that
are only slightly less conservative yield tighter constraints for
$R_{1.4}$, and a PDF with a smaller variance (the light-shaded red lines
refer to \cite{Drischler2016}); in particular, we obtain a variance of
$\sim 1\ \rm km$ at a $2$-$\sigma$ level on the radii of low-mass stars.
Figure~\ref{fig:new_EOSs} thus highlights that a more accurate knowledge
of the matter in the outer core, \ie for number densities in the range
$0.08\!\lesssim n/{\rm fm}^{-3}\!\lesssim0.21$, can have an enormous
impact on the macroscopic properties of neutron stars. Any progress in
this direction will impact our understanding of compact stars. For
completeness we comment that while the results of \cite{Drischler2017}
constrain the hadronic sector very well, we found their effect on the
twin-star branch to be less pronounced.

\noindent\emph{Conclusions.} Using a parameterized construction of the
EOS which matches realistic nuclear-physics calculations for the stellar
crust at very low densities and perturbative QCD calculations at very
high densities, we have constructed more than two million different EOSs,
with and without phase transitions, that are physically plausible and
compatible with the observations on the maximum mass. The corresponding
PDFs have been studied to set constraints on the plausible values for the
radii and tidal deformabilities of neutron stars. In particular, we have
studied how the PDFs are affected by the imposition of new constraints on
the maximum mass and on the tidal deformability that have been recently
deduced via GW170817. These additional constraints induce significant
changes in the PDFs, especially when they are imposed
simultaneously. While the statistical properties of the stellar models
vary only weakly with the maximum mass, constrains on the lower limit of
the tidal deformability exclude the softest EOSs and shift the peak of
the distribution to larger values, yielding a variance well-below two
kilometres for purely hadronic EOSs, \ie $12.00\!<\!R_{1.4}\!<\!13.45$ at
a $2$-$\sigma$ level, with a most likely value of
$\bar{R}_{1.4}=12.39$. On the other hand, the radii of twin stars are
less constrained, namely, $8.53\!<\!R_{1.4}\!<\!13.74$, with very compact
stars possible but not probable, so that $\bar{R}_{1.4}=13.06$. We have
also been able to set upper limits on the tidal deformability of hybrid
stars, \eg $\tilde{\Lambda}_{1.7}\!<\!461$. For hadronic EOSs an
additional tightening of the uncertainties is achieved with refined
descriptions of the outer core, thus calling for an improved
characterisation of the EOS at intermediate densities. Overall, our
results show that GW170817 has had a profound impact on our ability to
constrain the maximum mass, tidal deformability and radii of neutron
stars. New detections will provide even tighter constraints on the EOS of
nuclear matter and restrict the radius of neutron stars to below the
$10\%$ uncertainty
\cite{Rezzolla2016,Bose2017,Chatziioannou2017,Yang2017}.

\smallskip\noindent\emph{Acknowledgements.} It is a pleasure to thank
Christian Drischler, Arianna Carbone, Kai Hebeler, Jim Lattimer, Tyler Gorda
and Aleksi Vuorinen for useful discussions. 
Support comes from: ``PHAROS'', COST Action CA16214;
LOEWE-Program in HIC for FAIR; the DFG through the grant CRC-TR 211; the
European Union's Horizon 2020 Research and Innovation Programme (Grant
671698) (call FETHPC-1-2014, project ExaHyPE). ERM and LRW acknowledge
support from HGS-HIRe.


\bibliography{aeireferences,local}
\bibliographystyle{apsrev4-1}

\clearpage

\setcounter{figure}{0}

\section*{Supplemental material}
\subsection*{Parameterizing our ignorance} 
\label{sec:setup}

In the following we give a detailed description of the construction of
the EOS used in our work. In essence, for the low-density limit we use
the EOSs of Refs. \cite{Baym71b_2, Negele73_2}, that describe the neutron
star's crust up until a number density of $n_{\rm crust}=0.08\,{\rm
fm}^{-3}$. Instead, for baryon number densities $n_{_{\rm B}}$ found in
the outer core, \ie between $n_{\rm crust}$ and $n_{_{\rm
B}}\approx\!1.3\,n_{\rm sat}\!\approx\!  0.21\,{\rm fm}^{-3}$, where
$n_{\rm sat}=0.16\,{\rm fm}^{-3}$ is the nuclear-saturation density, we
use an improved and consistent neutron-matter EOS based on the chiral
expansion at ${\rm N}^3{\rm LO}$ of the nucleon-nucleon and three-nucleon
(3N) chiral interactions that takes into account also the subleading 3N
interactions, as well as 4N forces in the Hartree-Fock
approximation \cite{Drischler2016_2}. In practice, we take the lower and
upper limits of the uncertainty band for this EOS and fit them with two
polytropes. The first polytrope is used up until $\approx\!n_{\rm sat}$
and yields an adiabatic index $\Gamma_1$ in the range $[1.31,1.58]$,
while the second adiabatic index $\Gamma_2$ is chosen in the range
$[2.08,2.38]$; the limits of these ranges essentially establish the
uncertainty in the description of the EOS in the outer core. We note that
taking just these very stiff and soft limits into account -- as done by
Ref. \cite{Annala2017_2} -- only allows to give upper and lower bounds, as
set by the softest and stiffest EOS possible. On the other hand, to
understand common features and effects of the crust and outer core on the
neutron-star models, we construct our EOSs from uniform distributions of
$\Gamma_1$ and $\Gamma_2$, which will obviously include the stiffest and
softest possible EOSs, but also all the other possibilities between these
two limits. In addition, and to explore the sensitivity of our results on
this prescription of the EOS, we also investigate the impact of a further
refinement of the EOS obtained using a new many-body Monte-Carlo
framework for perturbative calculations applied to a set of chiral
interactions \cite{Drischler2017_2}. In this case, we use the range over six
EOSs, each based on a different Hamiltonian \cite{Hebeler2011_2}, as uncertainty band.
\begin{figure}[h!]
  \includegraphics[width=0.9\columnwidth]{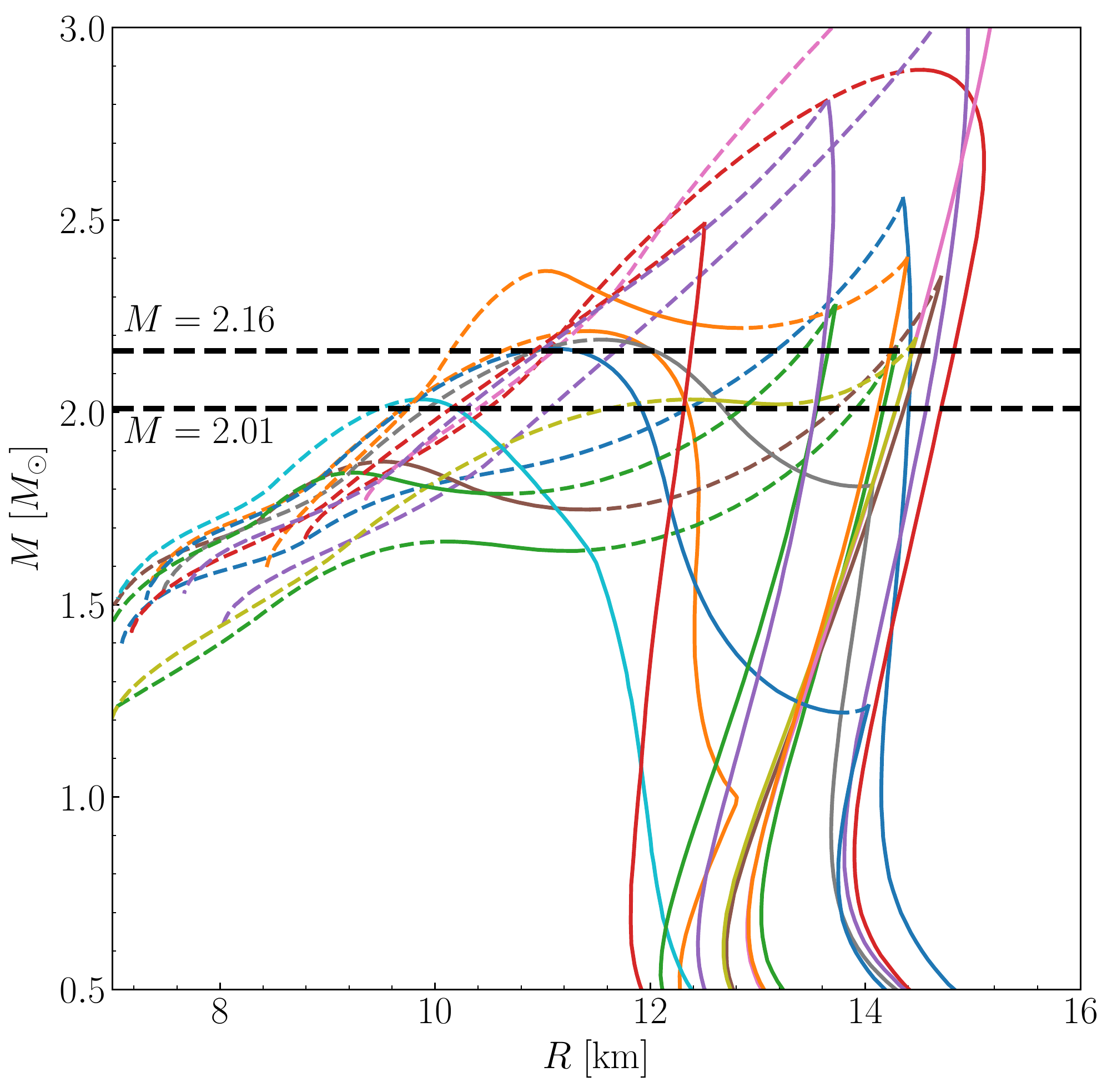}
  \caption{\footnotesize Representative sample of $(M,R)$ curves
    constructed from EOSs with a phase transition. While stable
    branches are shown as solid lines, the unstable ones are denoted by
    dashed ones. Also shown with thick dashed lines are the lower and
    upper constraint on the maximum mass.}
  \label{fig:sup_1}
\end{figure}

\begin{figure}[h!]
  \includegraphics[width=0.9\columnwidth]{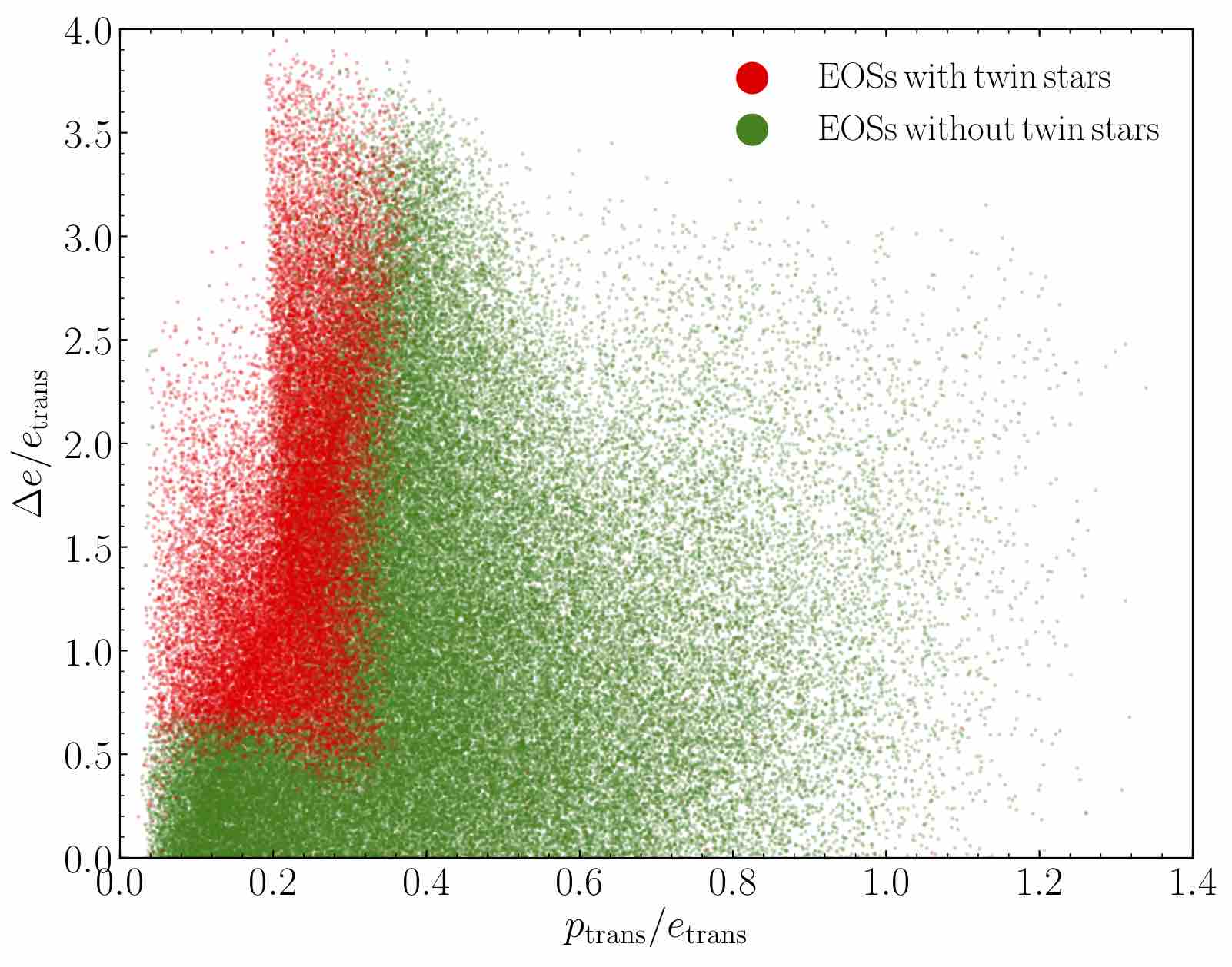}
  \caption{\footnotesize
 Relative jump of the energy discontinuity $\Delta e/e_{\rm trans}$ shown
  as a function of the normalised pressure at the phase transition
  $p_{\rm trans}/e_{\rm trans}$. All EOSs represented posses a phase
  transition but some lead to twin stars (red dots), while others do not
  (green dots).}  \label{fig:sup_2}

\end{figure}
\begin{figure*}[t!]
  \includegraphics[width=1.8\columnwidth]{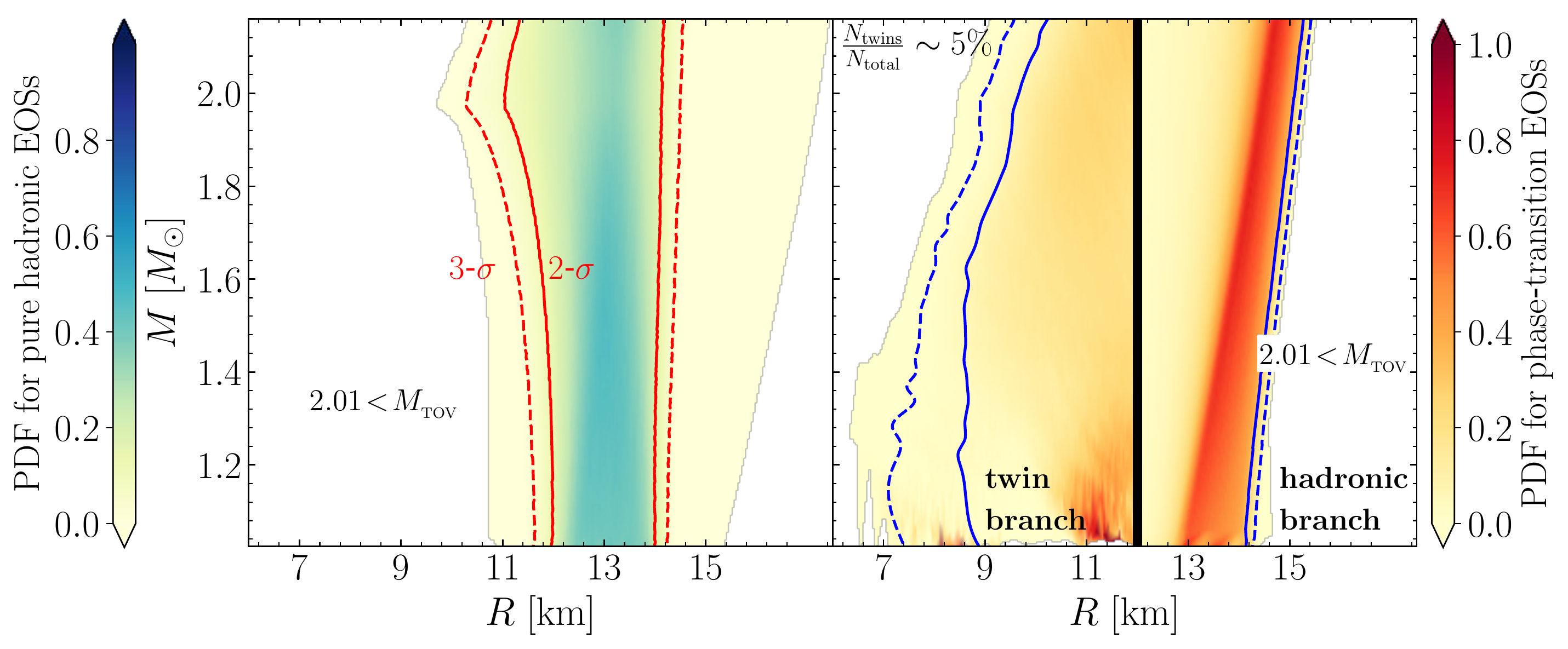}
  \caption{\footnotesize PDFs of stellar radii. Left panel: PDF with only
    the observational constraints on the observed maximum mass for pure
    hadronic EOSs; right: the same but for EOSs with a phase transition
    and where the PDF for $R<12 \, \rm km$ is for the twin-branch and the
    one for $R>12 \, \rm km$ for the hadronic branch.  In both panels the
    solid and dashed lines indicate the $2$-$\sigma$ and $3$-$\sigma$
    confidence levels, respectively. This figure should also be
    contrasted with Fig. 1 of the main text.}
  \label{fig:sup_3}
\end{figure*}

For the high-density part of the EOS, on the other hand, we follow
\cite{Annala2017_2} and \cite{Kurkela2014_2}, and use the cold quark-matter
EOS derived by \cite{Fraga2014_2}, which is based on the perturbative QCD
calculation of \cite{Kurkela2010_2}. The uncertainty in this EOS is
estimated by changing the renormalization scale parameter within a factor
two, $X\in[1,4]$, which is chosen from a uniform distribution, allowing
us to match the last segment of the interpolating piecewise polytrope via
its adiabatic index at the baryon chemical potential $\mu_{\rm
b}=2.6\,\rm GeV$. Finally, for the unconstrained region above $\sim
1.3\,n_{\rm sat}$, we follow \cite{Kurkela2014_2} and interpolate between
these limits using piecewise polytropes with four segments (tetratropes),
whose polytropic exponents, as well as the matching points between the
segments, are chosen at random assuming equal probabilities, ensuring
continuity of energy and pressure at the matching points. We have also
checked our results to be robust, when using five instead of four
polytropic segments.

After constructing an EOS with the procedure outlined above, we check
that it is physically plausible by ensuring that the sound speed is
subluminal everywhere inside the star, while the minimal thermodynamic
stability criteria is automatically satisfied. For any EOS passing this
test, we then construct a sequence of neutron-star models by solving the
TOV equations. Additionally, we check compatibility with observations of
the maximum mass of neutron stars \cite{Antoniadis2013_2} and ensure that
the maximum mass of any sequence exceeds $2.01$. In this way, we compute
a million of equilibrium sequences containing more than a billion stars
and essentially covering the whole possible range of EOSs from very soft
to very stiff.

As mentioned in the main text, in order to take into account the
possibility that the neutron-star matter has a strong phase transition in
its interior, we construct an equally sized set of $10^6$ EOSs using the
above setup, but adding a jump in energy density $\Delta e \in [0,1000]\,
{\rm MeV}/{\rm fm}^3$, while ensuring constancy of the chemical
potential. This jump is introduced randomly between the polytropic
segments and its range is motivated by the results of
Ref. \cite{Christian2018_2}. The latter choice ensures that all four
categories of twin stars can be obtained (see \cite{Christian2018_2} for an
overview). This is also evident from Fig. \ref{fig:sup_1}, which shows a
representative sample of the mass radius-curves obtained with the setup
for EOSs with phase transition. In particular, it is possible to see that
the phase transition can set in at very low masses, \ie $<\!1\,M_\odot$,
but also at very high masses, \ie $>\ 3\ M_\odot$, where the two stable
branches can be up to $5\ \rm km$ apart. Overall, we find that with this
prescription we can cover well the space of possible EOSs with a phase
transition.

A more detailed representation of the occurrence and properties of the
phase transitions in the EOSs constructed here can
be obtained by looking at Fig. \ref{fig:sup_2}, which shows the relative
length of the jump $\Delta e/e_{\rm trans}$ over $p_{\rm trans}/e_{\rm
trans}$, where $p_{\rm trans}$ is the pressure at the onset of the phase
transition and $e_{\rm trans}$ the corresponding energy. The figure
clearly shows that our sample of EOSs with phase transitions populates
the whole space necessary for obtaining all four categories of twin stars
(\cf Fig. 3 of Ref. \cite{Christian2018_2}). The red region also confirms that
twin stars can only be obtained for a small set of combinations of the
parameters $\Delta e$ and $e_{\rm trans}$. Additionally, we can see that
the remaining parameter space is also covered well.


\subsection{Comparing EOSs with and without phase transitions}

While purely hadronic EOSs yield neutron star-models with radii $\gtrsim 10\,
\rm km$, this is not contradicting the results of Ref.  \cite{Tews2018_2},
who also find models with radii as small as $\sim 8\,\rm km$. From Fig.
\ref{fig:sup_3}, where we show all models without any upper constraint on
the maximum mass and the tidal deformabilities, it is evident that such
small radii can only be obtained when a phase transition is present. At
the same time, it is evident that such compact stars are not likely and
require some fine-tuning of the EOS. For instance, we find that
such small radii are found for twin stars, whose total number, $N_{\rm
twins}$, only corresponds to $\sim 5\%$ of the total number of 
all models build from EOS exhibiting a phase transition, $N_{\rm total}$.
\begin{figure}[h!]
  \includegraphics[width=1.0\columnwidth]{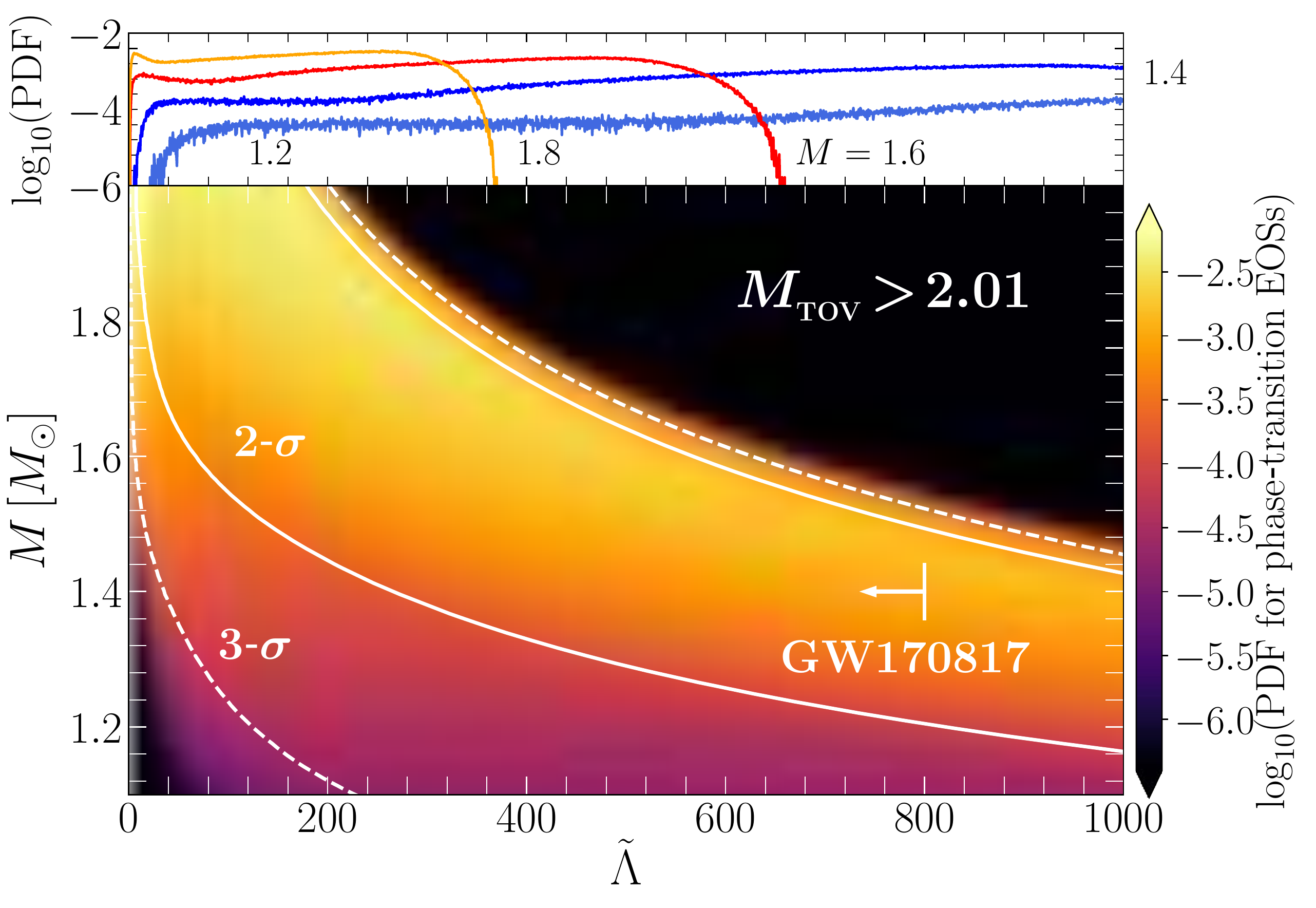}
  \caption{\footnotesize The same as Fig. 3 of the main text, but for
    the set of EOSs with phase transition.}
  \label{fig:sup_4}
\end{figure}

Another striking difference between purely hadronic EOSs and the ones
with phase transitions is the occurrence of small tidal deformabilities
in the latter case. For instance, for $M\lesssim 1.4$ models with
$\tilde{\Lambda}_{1.4} \lesssim 100$ can be found, as is evident from
Fig. \ref{fig:sup_4}, which is the same as Fig. 3 of the main text but
in the case of EOSs with a phase transition. This finding resolves the
apparent conflict between the small lower limit for
$\tilde{\Lambda}_{1.4}$ found in Ref. \cite{De2018_2} (\ie
$\tilde{\Lambda}_{1.4} > 75$) and the stricter one obtained by
Ref. \cite{Annala2017_2} (\ie $\tilde{\Lambda}_{1.4} > 120$) or the value
of $\tilde{\Lambda}_{1.4} > 375$ discussed in the main text. In the case
of Ref. \cite{De2018_2}, in fact, no distinction was made between purely
hadronic and EOSs with phase transitions.  On the other hand, a
comparison between Figs. 3 of the main text and \ref{fig:sup_4} clearly
shows that the lower limit on $\tilde{\Lambda}_{1.4}$ becomes much
stricter if one assumes a purely hadronic EOS. A similar conclusion has
been drawn in Ref. \cite{Tews2018_2}, where a model with phase transitions
($\tilde{\Lambda}_{1.4} > 80$) and one without ($\tilde{\Lambda}_{1.4} >
280$) was employed. In addition, we find that for $M \gtrsim 1.6$ a
strict cut-off for the upper limit of $\tilde{\Lambda}$ exists if a phase
transition is present. As a result, deducing a value of $\tilde{\Lambda}$
from a gravitational-wave measurement in this mass range could be used to
distinguish a purely hadronic EOS from one with a phase transition, as we
have further outlined in the main text.
%

%

\end{document}